\documentclass[aps,prmaterials,twocolumn,superscriptaddress]{revtex4-2}

\usepackage{mhchem}
\usepackage{bm}

\usepackage{graphicx}
\usepackage{caption}

\usepackage{siunitx}
\DeclareSIUnit\angstrom{\text {Å}}

\usepackage{jabbrv}
\usepackage{hyperref}
\hypersetup{colorlinks=true, citecolor=blue, urlcolor=blue, linkcolor=blue}

\begin{document}
\title{Theoretical antiferromagnetism of ordered face-centered cubic Cr-Ni alloys}

\author{Flynn Walsh}
\affiliation{Materials Sciences Division, Lawrence Berkeley National Laboratory, Berkeley, CA 94720}
\affiliation{Graduate Group in Applied Science \& Technology, University of California, Berkeley, CA 94720}
\author{Robert O. Ritchie}
\affiliation{Materials Sciences Division, Lawrence Berkeley National Laboratory, Berkeley, CA 94720}
\affiliation{Department of Materials Science \& Engineering, University of California, Berkeley, CA 94720}
\author{Mark Asta}
\email{mdasta@berkeley.edu}
\affiliation{Materials Sciences Division, Lawrence Berkeley National Laboratory, Berkeley, CA 94720}
\affiliation{Department of Materials Science \& Engineering, University of California, Berkeley, CA 94720}

\begin{abstract}
Contrary to prior calculations, the Ni-rich ordered structures of the Cr-Ni alloy system are found to be antiferromagnetic under semi-local density-functional theory.
The optimization of local magnetic moments significantly increases the driving force for the formation of \ce{CrNi2}, the only experimentally observed intermetallic phase.
This structure's \textit{ab initio} magnetism appears well described by a Heisenberg Hamiltonian with longitudinal spin fluctuations; itinerant Cr moments are induced only by the strength of exchange interactions.
The role of magnetism at temperature is less clear and several scenarios are considered based on a review of experimental literature, specifically a failure of the theory, the existence of an overlooked magnetic phase transition, and the coupling of antiferromagnetism to chemical ordering.
Implications for related commercial and high-entropy alloys are discussed for each case.
\end{abstract}

\maketitle

\section{Introduction} 

\begin{figure}
\includegraphics{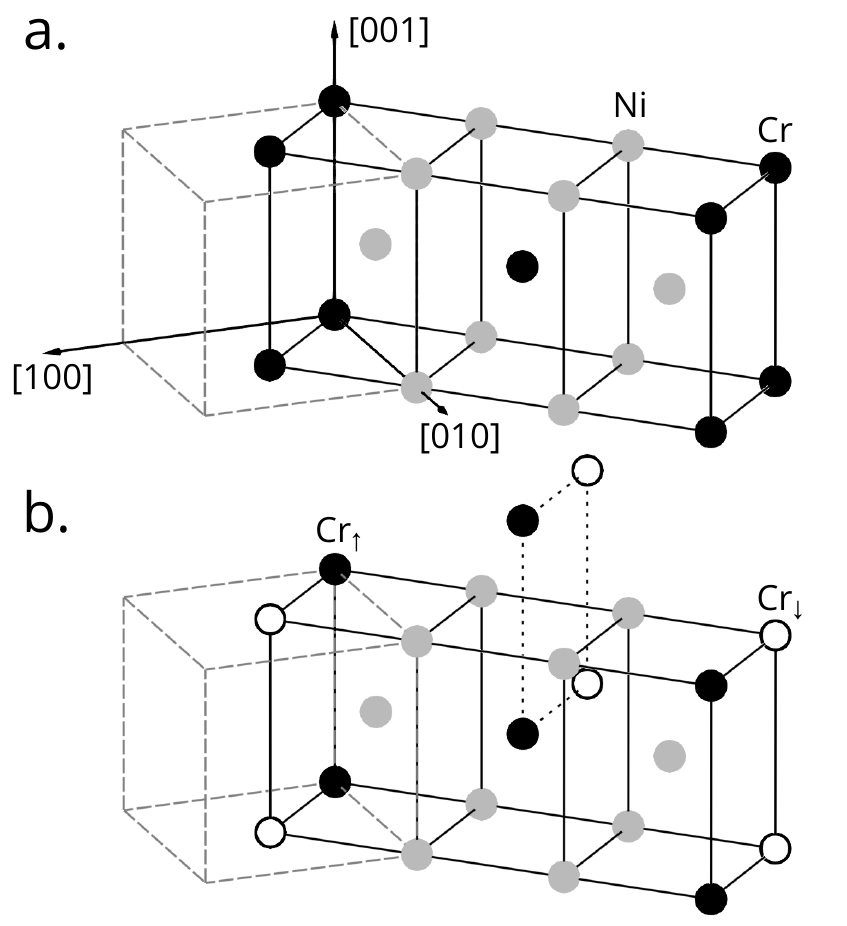}
\caption{\label{fig:crni2}
    (a) The conventional unit cell of \ce{CrNi2}. Dashed lines and coordinate axes indicate the conventional fcc unit cell.
    (b) Proposed AFM ground state of \ce{CrNi2}, drawn on the same structure.
    Arbitrarily oriented ``up" and ``down" Cr moments are respectively represented using $\bullet$ and $\circ$ markers. 
The actual tiling supercell is larger and an extension of the central $(\bar{1}10)$ plane is included to fully depict the ordering.
}
\end{figure}

Ni-rich Cr-Ni alloys have been studied for over a century on account of their extensive applications and intriguing process of chemical ordering.
While elemental Cr forms body-centered cubic (bcc) crystals, about \SI{36}{at{.}.\%} Cr is soluble in face-centered cubic (fcc) Ni.
\ce{CrNi2}, the system's only experimentally observed intermetallic phase, emerges from these solid solutions as a \ce{MoPt_2}-type lattice decoration---see Fig. \ref{fig:crni2}(a)---below \SI{\sim863}{K} \cite{nash86}.
This relatively low ordering temperature kinetically limits the realization of \ce{CrNi2}, which forms nanoscale antiphase domains that slowly grow over thousands of hours of annealing \cite{marucco88}.
Still, the structure has been observed in alloys with \SIrange{\sim25}{36}{at{.}.\%} Cr \cite{nash86,marucco95}, with indications of a similar phase found in commercial Ni-based alloys \cite{marucco88,park04} and fcc medium/high-entropy alloys \cite{jin17,du22} of current interest.

The gradual formation of \ce{CrNi2} is associated with well studied ``K-state" phenomena, including significant increases in hardness and resistivity, as well as lattice contraction \cite{taunt75,marucco88,marucco95}. 
Its magnetic properties, however, have largely escaped scrutiny.
While elemental Ni is ferromagnetic (FM) below \SI{631}{K} \cite{crangle71}, the addition of Cr rapidly decreases both spontaneous magnetization and Curie temperature, resulting in a ``quantum critical point" at about \SI{11}{at{.}.\%} Cr \cite{besnus72,vishvakarma21a}.
Alloys with greater concentrations of Cr are generally regarded as paramagnetic at all temperatures \cite{arajs67,hirabayashi69}, although evidence for a more nuanced picture will be discussed later. 

Regardless of finite temperature behavior, paramagnetism is not an electronic ground state and likely originates at least in part from thermally induced spin fluctuations, as have been theorized in chemically similar alloys \cite{ruban07,ruban16,dong17,dong21}. 
Many previous studies \cite{arya02,tucker08,wrobel15,goiri18,fedorov20}, including high-throughput databases \cite{curtarolo12,jain13}, have predicted that \ce{CrNi2} is nonmagnetic (NM) at \SI{0}{K}---given the apparent lack of magnetic order (and desire to model high-temperature conditions), others have neglected spin-polarization entirely \cite{rahaman14}.
However, while standard techniques can easily simulate ferromagnetism, the convergence of antiferromagnetic (AFM) electronic structures requires the input of a specific magnetic symmetry, including a commensurate simulation supercell.
(This is generally accomplished through qualitative initialization of local magnetic moments on atomic sites.)
Both the primitive and conventional unit cells of \ce{CrNi2} (see Fig. \ref{fig:crni2}) consider all nearest neighbor Cr pairs to be symmetrically equivalent, imposing identical magnetic states that preclude the realization of the simplest forms of antiferromagnetism.
It is not clear if any prior computational study of \ce{CrNi2} investigated supercells compatible with AFM order, let alone seeded reasonable magnetic structures, motivating a revisitation of magnetism in Cr-Ni alloys under conventional electronic structure methods (see Sec. \ref{sec:methods}).

Secs. \ref{sec:crni2} \& \ref{sec:crni3} detail how previously overlooked AFM states theoretically exist for ordered Cr-Ni structures across a range of compositions. 
The antiferromagnetism of \ce{CrNi2} is further characterized in Sec. \ref{sec:fit} in terms of a Heisenberg model with longitudinal spin fluctuations (LSFs).
While the nature of magnetism at finite temperature remains unclear, Sec. \ref{sec:neel} attempts to explain the ground state predictions in light of experimental findings, although it is concluded that further measurements are necessary.

\section{Methods}\label{sec:methods}

For Secs. \ref{sec:crni2} \& \ref{sec:crni3}, collinearly spin-polarized DFT structure optimizations were performed using the Vienna \textit{Ab initio} Simulation Package (VASP) \cite{kresse93}.
(None of the considered structures contained geometric frustration that could be expected to elicit noncollinear moments and spin-orbit coupling is negligible at the energy scale of ordering.)
Electronic states were represented in terms of plane-waves \cite{kresse96,kresse96a} with a \SI{520}{eV} cutoff and a linear $k$-point density of \SI{0.15}{\AA}.
State occupancies were smeared to a width of 0.1 eV using a first-order Methfessel-Paxton method \cite{methfessel89}. 
Atomic moments were conservatively integrated within \SI{1}{\angstrom} spheres so that all magnetization was contained within the Cr Bader surface \cite{tang09}---more rigorous methods for determining local moments, such as the complete integration of Bader volumes, were complicated by the subsequent imposition of spin-spirals.
Cr sites were assigned initial moments of $\pm\SI{2}{\mu_{\rm{B}}}$; Ni sites were also initialized with $\SI{1}{\mu_{\rm{B}}}$ moments, but expectedly demagnetized during the convergence of calculations unless specifically noted.

The treatment of electronic exchange and correlation (XC) requires some investigation.
Sec. \ref{sec:crni2} considers both the local spin-density approximation (LSDA) and the generalized-gradient approximation (GGA), the former using Perdew and Zunger's \cite{perdew81} parameterization of Ceperley and Alder's \cite{ceperley80} correlation energies and the latter according to Perdew, Burke, and Ernzerhof, either in the original formulation (PBE) \cite{perdew96} or as revised for solids (PBEsol) \cite{perdew08}.
Correlation energies were interpolated using the method of Vosko, Wilk, and Nusair \cite{vosko80}.
For calculations at the experiment lattice parameter, PBE structures were scaled to \SI{3.5240}{\angstrom} for Ni, \SI{2.8848}{\angstrom} for Cr \cite{haynes16}, and \SI{3.562}{\angstrom} for \ce{CrNi2} \cite{hirabayashi69}.

On the basis of the results presented in Sec. \ref{sec:crni2}, PBE was selected for all further calculations; this choice is largely motivated by results for elemental Cr, which require some further discussion. 
Namely, it must be noted that semi-local DFT fails to reproduce the experimental ground state of bcc Cr \cite{hafner02}, which is a structurally incommensurate spin-density wave described by the wavevector $\bm{q} \sim 0.95\bm{b}_1$ \cite{fawcett88}, where $\bm{b}_1$ is the cubic reciprocal lattice vector.
Still, the observed spin-density wave only slightly differs from the simple AFM structure predicted by DFT ($\bm{q} = \bm{b}_1$), corresponding to an energy difference of a few meV per atom in the present calculations.
Absent further failures, DFT may still reasonably approximate the magnetic ground state of Cr, which provides an experimental benchmark for assessing predictions of antiferromagnetism in Cr-Ni alloys.

In order to better characterize the magnetism of \ce{CrNi2}, spin-wave calculations were performed in Sec. \ref{sec:fit}.
In these simulations, noncollinear AFM structures were represented as plane-waves of local magnetization density lying within the first Brillouin zone of a structurally minimal unit cell \cite{sandratskii98}, as implemented in VASP.
\ce{CrNi2} was modeled using a static primitive lattice with $\bm{a}_1 = \frac{1}{2}[110]$, $\bm{a}_2= \frac{1}{2}[\bar{1}21]$, and $\bm{a}_3 = [001]$ and, in reciprocal space, $\bm{b}_1 = \frac{1}{3}[420]$, $\bm{b}_2 = \frac{1}{3}[\bar{2}20]$, and $\bm{b}_3 = \frac{1}{3}[1\bar{1}3]$. (The exact structure was determined from the magnetic ground state, restricting magnetovolume coupling and slightly exaggerating the magnetization energy, cf. Table \ref{table:afm}.)
Under this convention, the AFM ground state depicted in Fig. \ref{fig:crni2}(b) corresponds to a $[\frac{1}{2}00]$ wavevector, with additional structures determined from modulations of this state, e.g., AFM decoration along [001] is represented by the wavevector $[\frac{1}{2}0\frac{1}{2}]$.
The longitudinal degree of Cr magnetization was allowed to relax in all calculations.
Altogether, 111 wavevectors were chosen by interpolating among high-symmetry points near the region of AFM stability within the Brillouin zone.

\section{Results}
\subsection{Magnetic ground state of CrNi$_2$}\label{sec:crni2}

\begin{table}
    \caption{\label{table:afm} Several possible magnetic configurations of \ce{CrNi2} with AFM nearest $\left(\frac{1}{2}[110]\right)$ neighbors, according to several XC functionals.
    The first two columns indicate the type of magnetic ordering along the specified crystal direction. The third and fourth columns respectively give the average local moment magnitude and formation energy, while the last value is the change in energy relative to the NM structure.
Calculations for AFM bcc Cr and NM \ce{CrNi2} are provided for comparison.}
    \begin{ruledtabular}
        \begin{tabular}{cccccc}
            \rule[-1.25ex]{0pt}{0pt} $[001]$ & $\frac{1}{2}[\bar{3}30]$ & $\lvert m_{\text{Cr}} \rvert $ ($\mu_B$) & $E_{\text{form.}}$ $\big(\rm{\frac{meV}{atom}}\big)$ & $ E_{\text{mag.}}$ $\big(\rm{\frac{meV}{Cr}}\big)$\\
        \hline\hline\\[-2.2ex]
        \multicolumn{5}{c}{GGA-PBE/self-consistent lattice} \\
        \hline\\[-2.2ex]
            AFM & AFM & 1.16 & -26.7 & -24.1 \\
            AFM & FM & 1.20 & -28.5 & -29.4 \\
            FM & FM & 1.29 & -30.8 & -36.4 \\
            FM & AFM & 1.31 & -31.9 & -39.6 \\
            \hline\\[-2.2ex]
            \multicolumn{2}{c}{NM \ce{CrNi2}} & --- & -18.7 & --- \\
            \multicolumn{2}{c}{bcc Cr} & 0.98 & 0.0 & -15.9 \\
        \hline\hline\\[-2.2ex]
        \multicolumn{5}{c}{GGA-PBEsol/self-consistent lattice} \\
        \hline\\[-2.2ex]
            AFM & AFM & 0.45 & -41.7 & -1.0 \\
            AFM & FM & 0.56 & -42.3 & -2.7 \\
            FM & FM & 0.74 & -42.8 & -4.4 \\
            FM & AFM & 0.79 & -43.0 & -4.9 \\
            \hline\\[-2.2ex]
            \multicolumn{2}{c}{NM \ce{CrNi2}} & --- & -41.4 & --- \\
            \multicolumn{2}{c}{bcc Cr} & 0.34 & 0.0 & -0.6 \\
        \hline\hline\\[-2.2ex]
        \multicolumn{5}{c}{LSDA/experimental lattice constant} \\
        \hline\\[-2.2ex]
            AFM & AFM & 0.75 & -40.0 & -2.1 \\
            AFM & FM & 0.83 & -40.8 & -4.7 \\
            FM & FM & 0.92 & -41.9 & -7.9 \\
            FM & AFM & 0.97 & -42.4 & -9.3 \\
            \hline\\[-2.2ex]
            \multicolumn{2}{c}{NM \ce{CrNi2}} & --- & -39.3 & --- \\
            \multicolumn{2}{c}{bcc Cr}        & 0.64 & 0.0 & -2.5 \\
        \hline\hline\\[-2.2ex]
        \multicolumn{5}{c}{GGA-PBE/experimental lattice constant} \\
        \hline\\[-2.2ex]
            AFM & AFM & 1.30 & -25.4 & -41.8 \\
            AFM & FM & 1.33 & -27.4 & -47.9 \\
            FM & FM & 1.40 & -30.0 & -55.8 \\
            FM & AFM & 1.41 & -31.0 & -58.7 \\
            \hline\\[-2.2ex]
            \multicolumn{2}{c}{NM \ce{CrNi2}} & --- & -11.4 & --- \\
            \multicolumn{2}{c}{bcc Cr} & 1.09 & 0.0 & -28.3 \\
        \hline\hline\\[-2.2ex]
        \multicolumn{5}{c}{GGA-PBEsol/experimental lattice constant} \\
        \hline\\[-2.2ex]
            AFM & AFM & 1.17 & -31.3 & -25.6 \\
            AFM & FM & 1.21 & -33.1 & -30.8 \\
            FM & FM & 1.28 & -35.3 & -37.4 \\
            FM & AFM & 1.29 & -36.1 & -40.0 \\
            \hline\\[-2.2ex]
            \multicolumn{2}{c}{NM \ce{CrNi2}} & --- & -22.8 & --- \\
            \multicolumn{2}{c}{bcc Cr} & 0.97 & 0.0 & -17.8 \\
        \end{tabular}
    \end{ruledtabular}
\end{table}

Metallic magnetism is typically understood in terms of two-site exchange couplings governed by Ruderman-Kittel-Kasuya-Yosida (RKKY) interactions, which decay according to the third power of distance in the long-range limit \cite{blundell01,turek06}.
Assuming AFM alignment of nearest neighbor Cr spins, the \ce{MoPt2}-type structure of \ce{CrNi2} can accommodate several distinct magnetic orderings depending on the nature of longer range exchange interactions.
As depicted in Fig. \ref{fig:crni2}(a), Cr atoms occupy every third plane in the $(\bar{1}10)$ direction; these planes are offset such that, if nearest neighboring Cr are AFM, every Cr-Cr bond between adjacent planes is balanced by an equidistant, opposite-spin counterpart (e.g., $\frac{1}{2}[\bar{1}21]$ and $\frac{1}{2}[\bar{2}11]$), preventing pairwise interactions between immediately neighboring planes from affecting the magnetic ground state.
It is thus assumed that the type of exchange interaction between second-nearest $[001]$ neighbors fixes the ordering of a given $(\bar{1}10)$ plane.
The relative order of the next-nearest $(\bar{1}10)$ planes then allows two possible magnetic structures for a given planar ordering, corresponding to either FM or AFM coupling between tenth-nearest $\frac{1}{2}[\bar{3}30]$ neighbors.

Altogether, these considerations allow four unique AFM structures, which are described in the first two columns of Table \ref{table:afm} in terms of $[001]$ and $\frac{1}{2}[\bar{3}30]$ exchange couplings.
The local moments and energies of these structures were calculated using several XC functionals and are tabulated in subsequent columns.
Both formation energies ($E_{\rm{form.}}$), relative to AFM bcc Cr and FM fcc Ni, and magnetization energies ($E_{\rm{mag.}}$), relative to NM structures, are given; the latter are normalized per Cr atom as Ni sites are NM under all of the considered scenarios.
Calculations for NM \ce{CrNi2} and elemental Cr are also provided.

Under the LSDA, the equilibrium lattice of \ce{CrNi2} is NM, but this is hardly surprising as the theory also fails to self-consistently reproduce the antiferromagnetism of bcc Cr, to say nothing of its inaccuracy for Ni.
Imposing the experimental lattice constant stabilizes magnetic order in both Cr and \ce{CrNi2}, although, at least in the former case, the magnetization energy is unphysically small \cite{hafner02}.
The calculated formation energies, on the other hand, are somewhat larger than expected from the relatively low experimental order-disorder transition temperature \cite{rahaman14,barnard14}.
In contrast, the PBE-based calculations predict a reasonable magnetization energy for Cr and even stronger magnetism in \ce{CrNi2}.
(It should be noted that the PBE local moments of bcc Cr appear somewhat larger than experiment \cite{fawcett94,hafner02}, although assigning AFM moments is somewhat less straightforward than determining FM magnetization.)
Using PBEsol leads to unrealistically weak antiferromagnetism in Cr, although the magnetization energy of \ce{CrNi2} remains several times larger and PBEsol calculations at the experimental lattice constant resemble equilibrium PBE.

Regardless of the specific XC functional, the relative hierarchy of magnetic interactions is clear: $[001]$ neighbors align ferromagnetically, as expected from RKKY theory, while interplanar $\frac{1}{2}[\bar{3}30]$ neighbors slightly favor AFM coupling, as depicted in Fig. \ref{fig:crni2}(b).
Moreover, in all physically plausible scenarios in which Cr is correctly AFM, the magnetization energy of \ce{CrNi2} is more than twice that of bcc Cr on a per Cr basis; the local moments of \ce{CrNi2} are also consistently larger.
Quantitatively, PBE clearly provides the most reasonable description of Cr and is used for the remainder of the study.
While PBE produces the largest absolute magnetization energies, the predicted the ratio of $E_{\rm{mag.}}^{\ce{CrNi2}}$ to $E_{\rm{mag.}}^{\rm{Cr}}$ is comparable to or significantly less than other calculation schemes.
Its formation energies also seem reasonable \cite{rahaman14,barnard14}, although these values are notably affected by magnetic order; the formation energy of the optimized magnetic structure (\SI{-31.9}{meV/atom}) is seventy percent larger in magnitude than the NM equivalent (\SI{-18.7}{meV/atom}).
All PBE formation energies are plotted in Fig. \ref{fig:hull}.

\subsection{Magnetism of other fcc orderings}\label{sec:crni3}

\begin{figure}
\includegraphics{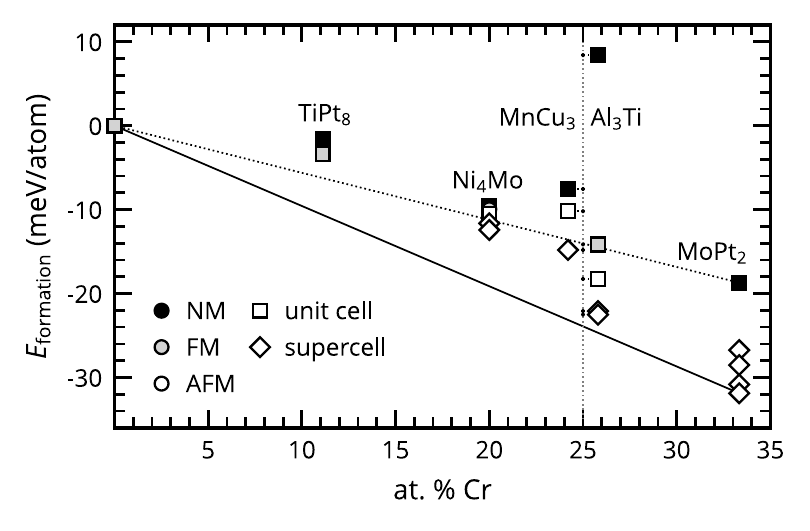}
\caption{
\label{fig:hull}
Calculated formation energies of various orderings discussed in the text as a function of composition.
Annotations provide the prototype of each structure, with markers indicating the type of simulation cell and converged magnetic order.
(A ``unit cell" may be primitive or conventional.)
The dotted and solid lines represent the convex hulls of stable structures based, respectively, on nonmagnetic and magnetic \ce{CrNi2}.
}
\end{figure}
\begin{figure}
\includegraphics{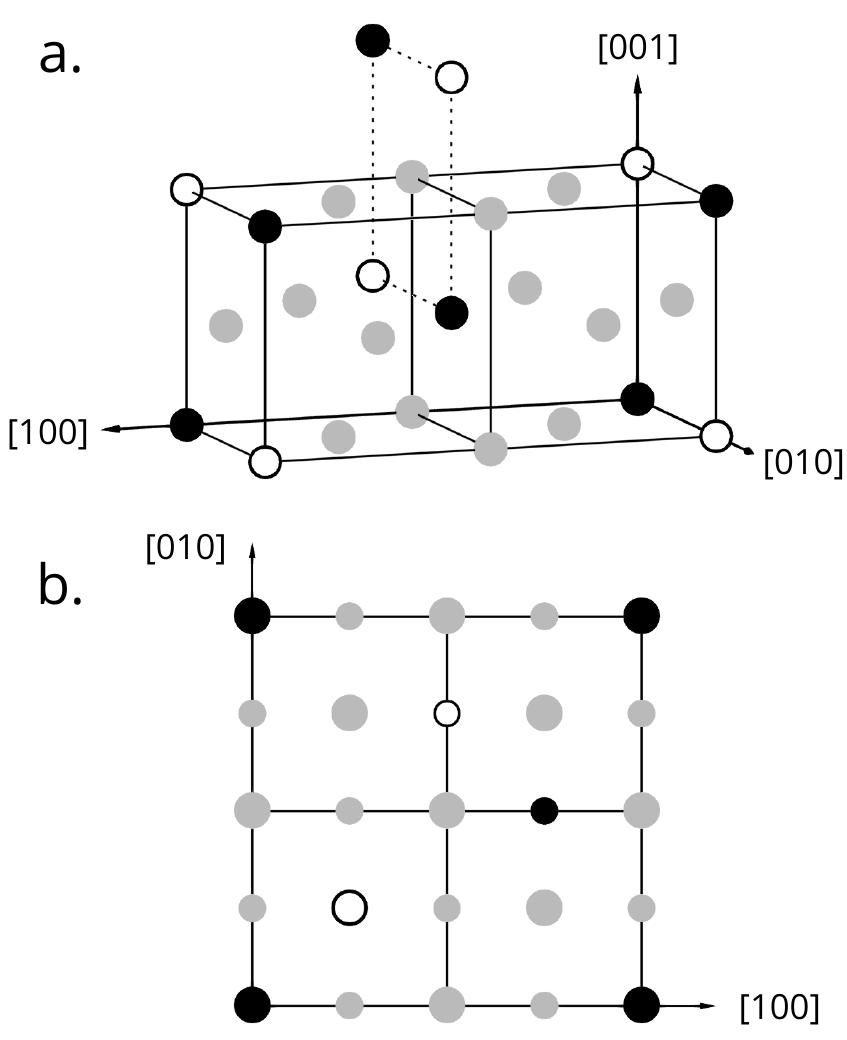}
\caption{\label{fig:crni3}
    Proposed AFM ground states for the (a) \ce{Al3Ti}-type and (b) \ce{MnCu3}-type orderings of \ce{CrNi3}, illustrated in the manner of Fig. \ref{fig:crni2}(b).
    The perspective of (b) is along $[001]$, in which Cr moments are AFM.
    Atomic sites are drawn with different sizes to illustrate alternating $(001)$ and $(002)$ planes.
    While \ce{MnCu3}-order is electronically more favorable, exchange interactions render \ce{Al3Ti}-type order lower energy, although neither is expected to be stable. 
}
\end{figure}

While only \ce{MoPt2}-type ordering has been observed in fcc Cr-Ni, the possibility of additional phases has received some prior attention. 
Regardless of composition, the system's high-temperature chemical short-range order (CSRO) maximizes diffuse scattering intensity at the $[\frac{1}{2}10]$ special point in reciprocal space rather than the $[\frac{2}{3}\frac{2}{3}0]$ point expected for \ce{MoPt2} \cite{schweika88,schonfeld88,caudron92,schonfeld94}, as can be explained by nucleation considerations \cite{defontaine81}.
Maxima at $[\frac{1}{2}10]$ are instead associated with the \ce{NiMo} (\ce{A2B2}), \ce{Al3Ti} ($\rm{D0_{22}}$/\ce{A3B}), and \ce{MnCu3} ($\rm{D0_{60}}$) \cite{rahaman14} prototypes, which have been considered by a number of previous studies. 

As for \ce{CrNi2}, the AFM ground states of other structures appear to have been overlooked with significant energetic consequences.
For example, the NM calculations of Ref. \cite{rahaman14} suggest that \ce{MnCu3} and \ce{Ni4Mo}-type ($\rm{D1_a}$) orderings could occur at low temperatures.
The introduction of antiferromagnetism, however, not only further stabilizes \ce{CrNi2}, but also inverts the relative favorability of \ce{MnCu3} and otherwise high-energy \ce{Al3Ti}, as shown in Fig. \ref{fig:hull} (relative to AFM bcc Cr and FM fcc Ni).
Fig. \ref{fig:crni3} depicts the AFM ground states predicted for the (a) \ce{Al3Ti} and (b) \ce{MnCu3} prototypes, which were identified through a similar process as for \ce{CrNi2}.
AFM decorations of the \ce{Ni4Mo}-type structure were also examined, although with minimal energetic effect. 

Ref. \cite{fedorov20} additionally calculated that a \ce{TiPt8}-type \cite{pietrokowsky65} ordering of \ce{CrNi8} was stable relative to NM \ce{CrNi2} and FM Ni.
At this composition, Ni atoms retain FM magnetization that induces opposite moments in distantly spaced Cr sites, leaving little opportunity for other forms of antiferromagnetism.
As shown in Fig. \ref{fig:hull}, this structure was found to be slightly higher energy than in Ref. \cite{fedorov20}, although the disagreement is within the range expected from differences in simulation parameters.

\subsection{Magnetic parameterization of \ce{CrNi2}}\label{sec:fit}
\begin{figure}
\includegraphics{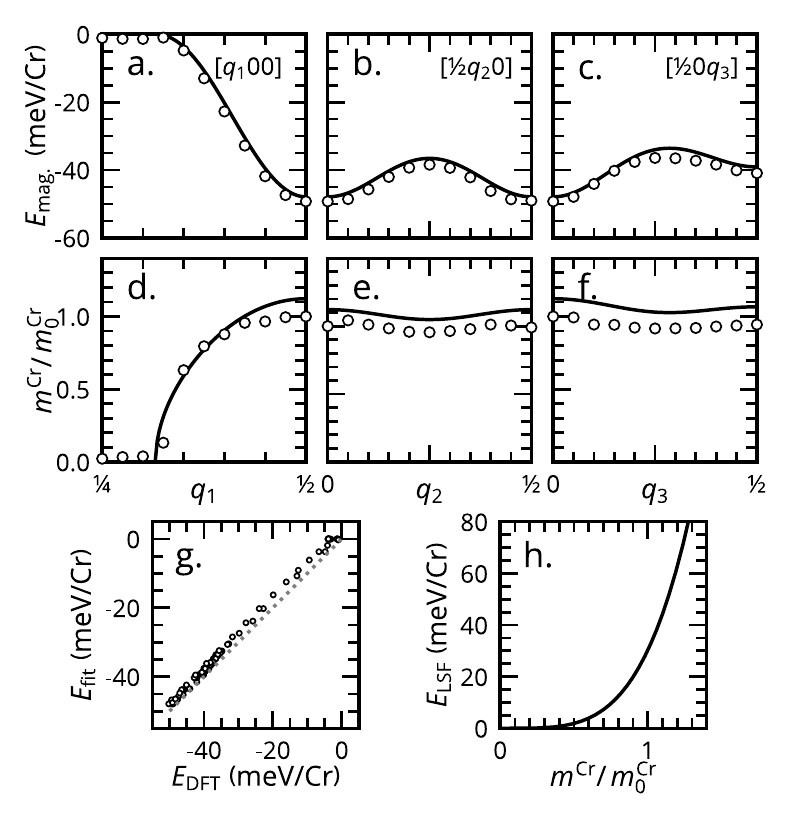}
\caption{\label{fig:fit}
    (a-c) \textit{Open circles}: DFT energies of spin spirals representing modulations of the AFM ground state along the reciprocal lattice vectors of \ce{CrNi2}, as indicated with fractional coordinates---the structure in Fig. \ref{fig:crni2}(b) corresponds to $[q_1 q_2 q_3] = \left[\frac{1}{2}00\right]$. 
    \textit{Solid lines}: equivalent values determined using Eq. (\ref{eq:H}) and Table \ref{table:Jij}. 
    (d-f) As above, relaxed Cr moments from the same calculations and minimum energy values according to the effective Hamiltonian.
    (g) The energies of all considered spin-spirals, as the determined from the Hamiltonian vs. as calculated with DFT.
    (h) Energy contribution from the longitudinal term of Eq. (\ref{eq:H}) as a function of local moment.
}
\end{figure}
\begin{table}
    \caption{\label{table:Jij} Effective Cr-Cr exchange parameters of Eq. (\ref{eq:H}) fit to DFT calculations for spin-spirals in ordered \ce{CrNi2}.
    The longitudinal energy is fully accounted for by $J_2 = \SI{31.95}{meV}$.
    Neighbor directions ($\bm{R}_{ij}$) are expressed in terms of the fcc lattice constant ($a$) for the structure depicted in Fig. \ref{fig:crni2}.
    The square of each neighbor distance is provided to simplify notation.}
    \begin{ruledtabular}
        \begin{tabular}{rccccc}
            \rule[-1.25ex]{0pt}{0pt} 
            $\bm{R}_{ij}$ ($a$) & $R_{ij}^2$ ($a^2$) & degeneracy & $J_{ij}$ (meV) \\
        \hline
            \rule{0pt}{2.5ex}
            $\frac{1}{2}[110]$ & 1/2 & 2 & -18.6 \\
            $[001]$ & 1 & 2 & 0.71 \\
            $\frac{1}{2}[112]$ & 3/2 & 4 & -0.36 \\
            $\frac{1}{2}[\bar{1}21]$ & 3/2 & 8 & --- \\
            $[110]$ & 2 & 2 & 14.91 \\
            $\frac{1}{2}[\bar{3}01]$ & 5/2 & 8 & --- \\
            $[111]$ & 3 & 4 & --- \\
            $\frac{1}{2}[\bar{1}23]$ & 7/2 & 8 & --- \\
            $[002]$ & 4 & 2 & --- \\
            $\frac{1}{2}[\bar{3}30]$ & 9/2 & 4 & -0.45 \\
            $\frac{1}{2}[033]$ & 9/2 & 8 & --- \\
            $\frac{1}{2}[114]$ & 9/2 & 4 & -0.57 \\
            $\frac{1}{2}[141]$ & 9/2 & 8 & 1.43 \\
            $[\bar{1}20]$ & 5 & 4 & --- \\
            $\frac{1}{2}[332]$ & 11/2 & 4 & --- \\
            $\frac{1}{2}[\bar{3}32]$ & 11/2 & 4 & -0.99 \\
        \end{tabular}
    \end{ruledtabular}
\end{table}

The introduction of AFM order further stabilizes \ce{CrNi2}, confirming the unique importance of this phase.
Its theoretical magnetic structure can be more completely characterized in terms of a Heisenberg model of exchange interactions.
The inconstant local moment recorded in Table \ref{table:afm}, as well as instability of FM \ce{CrNi_2}, indicates that the magnitude of Cr local moment is an important variable, as previously noted in austenitic stainless steels \cite{ruban16}. 
Changes in energy associated with such longitudinal spin fluctuations (LSFs) were accounted for by a phenomenological Landau-type expression, i.e. an energetic term proportional to the first few even powers of per site local moment \cite{murata72,uhl96,rosengaard97,dudarev07}.
Given the subjectivity inherent in localizing AFM moments, it is helpful to define a dimensionless effective spin $\bm{\mathcal{S}} = \bm{m} / m_0$, where $\bm{m}$ are classical magnetic moments (i.e., quantum expectation values) computed from density-functional theory (DFT) and $m_0$ refers to the ground state structure (\SI{1.31}{\mu_B} per Table \ref{table:afm}). 
For Cr sites indexed by $i$ and $j$, the model Hamiltonian takes the form
\begin{equation}\label{eq:H}
    H = \sum_i \sum_{p=1}^{p_{\rm{max}}} J_p \, \bm{\mathcal{S}}_i^{2p} - \sum_{i,j} J_{ij} \, \bm{\mathcal{S}}_i \cdot \bm{\mathcal{S}}_j
\end{equation}
where $J_{p}$ parameterize single-site LSFs and $J_{ij}$ describe exchange between Cr atoms at sites $i$ and $j$.

$J_p$ and $J_{ij}$ were fit to reproduce the energies of magnetic structures represented by spin spirals, which can be efficiently computed using Bloch's theorem---see Sec. \ref{sec:methods} for details.
As an illustrative example, Fig. \ref{fig:fit} shows the energies (a-c) and local moments (d-f) of spirals representing modulations of the ground state along the reciprocal lattice vectors of the primitive unit cell.

Consideration of the sixteen nearest Cr-Cr interactions, which are described in the first three columns of Table \ref{table:Jij}, was required to reasonably represent all the magnetic configurations, although only eight distinct instances of $J_{ij}$ were determined to be nonzero. 
The AFM $\frac{1}{2}[110]$ interaction is expectedly the largest, although, interestingly, FM coupling between $[110]$ neighbors in the same direction appears nearly as strong.
All other exchange pairs are individually weaker by an order of magnitude, although are more frequent (see column 3) and still significantly affect energies. 
The LSFs of Cr appear well described by a single $\bm{\mathcal{S}}^4$ term with $J_2 = \SI{31.95}{meV}$, which is plotted in Fig. \ref{fig:fit}(h).
For spin-Hamiltonian calculations, the optimum local moment of a given spin-wave was analytically determined from the fitted parameters.
Both DFT energies and magnetic moments are replicated by the model, as shown in Fig. \ref{fig:fit}---since local moments are at no point explicitly fitted, faithful reproduction of these values suggests that the relevant physics are largely captured.  

The exchange parameters calculated for \ce{CrNi2} are at least qualitatively applicable to the other considered structures, the magnetization energies of which can be largely explained by the frequency of the two strongest Cr-Cr exchange pairs, $\frac{1}{2}[110]$ and $[110]$.
In the \ce{MnCu3} structure depicted in Fig. \ref{fig:crni3}(b), for instance, Cr have one $\frac{1}{2}[110]$ and zero $[110]$ Cr neighbors and consequently exhibit weak AFM coupling.
In contrast, the $[001]$ planar ordering of Cr in \ce{Al3Ti} provides four $[110]$ Cr neighbors per Cr atom, effecting a several times larger magnetic ordering energy.
Lacking Cr nearest neighbors, the \ce{Al3Ti} structure also accommodates a previously theorized \cite{jain13} FM structure that abuts the convex hull of NM \ce{CrNi2} in Fig. \ref{fig:hull}; this result illustrates how the partial consideration of spin-polarization can be misleading even if, largely by happenstance, the qualitative picture of ground state phase stability in Cr-Ni remains unchanged.

Unfortunately, this simple model is likely not sophisticated enough to quantitatively predict the finite temperature behavior of \ce{CrNi2}, even if the DFT results are largely physical.
While the fitted parameterization accurately describes periodic deviations the ground state, the effective exchange parameters of less ordered configurations can greatly differ \cite{ruban04} and Hamiltonians fit exclusively to ordered configurations are known to poorly estimate N\'eel temperatures \cite{ruban07}.
It is also less obvious how to model the majority element of Ni, which is NM in all ground state calculations, but may play a significant role at temperature due to LSFs, if to an unclear end.
Perfectly disordered Ni moments should, on average, contribute zero net exchange to Cr sites, although it is easy to imagine random fluctuations locally destabilizing AFM order.
Further, simulations of highly itinerant Cr moments should be validated in the pure element and we were unable to construct an equivalent model that could satisfactorily describe the antiferromagnetism of bcc Cr.
(This is not entirely surprising, as the nominal disordering of bcc Cr hardly resembles a classical phase transition \cite{kormann13}.)

\section{Discussion}\label{sec:neel}

Even without finite temperature calculations, the basic prediction of antiferromagnetism in \ce{CrNi2} can be assessed in light of existing experimental data.
If \ce{CrNi2} is not in fact AFM at \SI{0}{K}, the calculations of Sec. \ref{sec:crni2} would represent a fairly spectacular failure of a theory that otherwise reasonably describes the magnetism of $3d$ transition metals and would be an important finding in of itself.
Such an error could originate from an overestimation of exchange interactions by the GGA, which has been previously postulated for Cr \cite{singh92,hafner02}.
Sill, even if the employed methods involved significant systematic error, all calculations found that the magnetic subsystem of \ce{CrNi2} was significantly stronger than that of pure Cr.
Given that bcc Cr has a nominal N\'eel temperature of \SI{311}{K} \cite{fawcett94}, it seems plausible that magnetic order in \ce{CrNi2} could persist well above ambient temperatures---naively scaling $T_{\rm{N}}^{\ce{Cr}}$ according to the ratio of PBE magnetization energies would suggest $T_{\rm{N}}^{\ce{CrNi2}} \sim \SI{775}{K}$.

There has been some recent interest in the role of magnetic interactions---particularly those of Cr---in the chemical ordering of both steels \cite{wrobel15,ruban16} and medium/high-entropy alloys \cite{niu15, lavrentiev16, schonfeld19, fedorov20, walsh21}, largely on the basis of DFT computations that have been called into question by some \cite{sales16,yin20}.
(This is to say nothing of how disordered moments can affect properties such as stacking fault energies at finite temperatures \cite{vitos06,dong19,dong21}.)
\ce{CrNi2} should provide a representative, experimentally accessible benchmark to test theoretical predictions concerning the role of magnetism in transition metal alloys that are not simple ferromagnets.

If \ce{CrNi2} is indeed AFM at \SI{0}{K}, the corresponding magnetic order-disorder transition should be detectable under calorimetry and magnetometry, although the N\'eel temperature is very much uncertain.
Unfortunately, the experimental characterization of ordered \ce{CrNi2} is incomplete---the fully formed intermetallic phase is of little practical interest given the deleterious effects of ordering on mechanical properties, to say nothing of its sluggish formation.
Perhaps most relevantly, Ref. \cite{hirabayashi69} measured the heat capacity of samples that were quenched after annealing at \SI{773}{K} for up to \SI{2900}{hours}.
They reported specific heat values from 573 to \SI{1073}{K}, observing a single peak at the \SI{\sim863}{K} chemical order-disorder transition that increased in magnitude with annealing; Refs. \cite{masumoto54,vintaykin72,jin17} provide similar results within this temperature range.
The absence of lower temperature calorimetry seemingly allows the possibility that a N\'eel transition below \SI{\sim550}{K} has simply escaped detection.
While the oversight of a magnetic transition in such a well studied system would be surprising, it does not seem out of the question, particularly if antiferromagnetism is restricted to the fully ordered phase.
Alternatively, it is interesting to consider the possibility of AFM order coupling to the well known chemical order-disorder transition at \SI{\sim863}{K}.

Limited experimental data actually support the possibility of AFM order in Cr-Ni alloys, which was in fact first theorized by Ref. \cite{gomonkov62} on the basis of neutron scattering in nominally disordered alloys with \SI{5.98}{at{.}.\%} and \SI{8.26}{at{.}.\%} Cr.
At these compositions, the observed magnetism was interpreted as Cr moments collectively aligning antiparallel to otherwise FM Ni.
These measurements are corroborated by the observation of heat capacity peaks in \SI{6.72}{} and \SI{8.94}{at{.}.\%} alloys at, respectively, \SI{492}{K} and \SI{675}{K} \cite{alizade71}. 
The temperature and magnitude of the these features grows with Cr concentration, presumably merging with the aforementioned structural transition occurring below \SI{863}{K} in alloys with \SI{>20}{at{.}.\%} Cr \cite{jin17}, as is widely understood to represent chemical disordering.
Indeed, Ref. \cite{nash86} attributed the anomalies of Ref. \cite{alizade71} to a CSRO transition, although \SI{492}{K} (\SI{219}{\celsius}) would be an extremely low temperature to detect chemical rearrangement given the experimental heating rate of \SI{100}{K/hour}.
Moreover, the specific heat curve of Ref. \cite{alizade71} appears to diverge in a manner that is far more consistent with long-range magnetic ordering than CSRO.
If Ref. \cite{alizade71} indeed detected antiferromagnetism at low Cr concentrations, it seems likely that some form of magnetic order persists up to the chemical order-disorder transition in \ce{CrNi2}.

Ref. \cite{chechernikov64} appears to further corroborate this picture of magnetism, reporting susceptibility features consistent with an AFM transition in the range of \SIrange{200}{300}{K} for alloys with \SI{8.75}{at{.}.\%} and \SI{11.1}{at{.}.\%} Cr.
After \SI{8}{hours} of annealing at \SI{900}{K}, the presumed development of CSRO raised the apparent magnetic transitions to \SIrange{500}{600}{K}.
Curiously, Ref. \cite{arajs67} failed to reproduce the observations of Ref. \cite{chechernikov64}, but instead found anomalies in the magnetic susceptibility of alloys with 16.6, 22.0, and \SI{25.0}{at{.}.\%} Cr in the vicinity of the chemical ordering temperature.
Ref. \cite{arajs67} in fact attributed these features to $\ce{CrNi2}$, although it is unclear if the measurements indirectly reflect a CSRO transformation or possibly reveal an explicitly magnetic transition. 
In related commercial alloys, the formation of $\ce{CrNi2}$-based phases has also been noted to affect magnetic properties, although an exact mechanism has not been proposed \cite{park04,mamiya16}.

If antiferromagnetism persists to high temperatures, its neglect would be expected to cause errors in prior ordering theory.  
For instance, Refs. \cite{caudron92,schonfeld94} derived pair potentials from scattering experiments that notably underestimated the chemical ordering temperature of \ce{CrNi2}, suggesting neglected interactions and inviting speculation as to the missing physics.
However, lattice models fitted to NM \cite{rahaman14} or mostly NM \cite{barnard14} DFT calculations slightly \textit{overestimate} the ordering temperature of \ce{CrNi2} after correcting incomplete pair potentials with many-body interactions. 
Barring a fortuitous cancellation of errors, the relative accuracy of these calculations implies that magnetism plays a negligible role in the ordering of \ce{CrNi2}.
Still, it is worth noting that while the model of Ref. \cite{rahaman14} largely reproduces the CSRO of Ref. \cite{schonfeld88}, which examined samples that were quenched after equilibration at \SI{828}{K}, it appears to overestimate the equivalent \textit{in situ} measurements at \SI{993}{K} (\SI{25}{at{.}.\%} Cr) and \SI{1073}{K} (\SI{33}{at{.}.\%} Cr) \cite{caudron92}.
If, very speculatively, magnetic order affected the CSRO of quenched samples and Ref. \cite{rahaman14} overestimated electronic interactions in a manner compensating the omission of magnetism, the apparent discrepancy could be explained. 

\section{Summary \& Conclusion}

A complete treatment of magnetic order greatly affects the ground state energetics of Cr-Ni alloys under standard DFT, greatly increasing the stability of \ce{CrNi2}, which remains the only predicted ordered phase.
The role of magnetism at temperature is less clear, with three plausible scenarios.
In the first, the prediction of antiferromagnetism is simply erroneous and the application of the theory to similar systems should be re-examined.
Alternatively, \ce{CrNi2} could form an AFM structure at \SI{0}{K} that disorders below \SI{\sim550}{K} with minimal impact on chemical bonding, although it would still be interesting to assess if AFM order existed at ambient conditions. 
Most intriguingly, experimental literature appears to offer the possibility of a magnetic phase transition coupling to the chemical order-disorder transition at \SI{863}{K}.
Ultimately, further thermodynamic and magnetic measurements are needed to determine the nature and role of magnetism in ordered \ce{CrNi2}, with additional study likely required for less ordered alloys of great practical interest.
\\
\\
\\
\begin{acknowledgments}
This work was supported by the US Department of Energy, Office of Basic Energy Sciences, Materials Sciences and Engineering Division under contract No. DE-AC02-05CH11231 as part of the Damage-Tolerance in Structural Materials (KC13) program.
Simulations were performed using the Lawrencium computational cluster provided by the IT Division of Lawrence Berkeley National Laboratory (supported by the same office and contract number), as well as award No. BES-ERCAP0021088 of the National Energy Research Scientific Computing Center, a US Department of Energy Office of Science User Facility operated under the same contract number.
\end{acknowledgments}

\bibliographystyle{jabbrv_apsrev4-2}
\bibliography{main}

\end{document}